\documentclass[preprint,12pt]{elsarticle}
\usepackage{amsmath}
\usepackage{amssymb}
\usepackage[hyphens]{url}  % DO NOT CHANGE THIS
\usepackage{graphicx} % DO NOT CHANGE THIS
\usepackage{natbib}  % DO NOT CHANGE THIS AND DO NOT ADD ANY OPTIONS TO IT
\usepackage{caption} % DO NOT CHANGE THIS AND DO NOT ADD ANY OPTIONS TO IT
\usepackage{algorithm}
\usepackage{algorithmic}

\usepackage{newfloat}
\usepackage{listings}
\DeclareCaptionStyle{ruled}{labelfont=normalfont,labelsep=colon,strut=off} % DO NOT CHANGE THIS
\floatstyle{ruled}
\newfloat{listing}{tb}{lst}{}
\floatname{listing}{Listing}

\usepackage{booktabs}

\title{PAGE-RAG: Evidence-Grounded Adaptive Graph Retrieval for Long-Document Question Answering}

\author[inst1]{Xingyu Chen}
\author[inst2]{Junxiu An*}
\author[inst3,inst5]{Jun Guo}
\author[inst4]{Li Wang}

\affiliation[inst1]{organization={School of Software Engineering, Chengdu University of Information Technology}, city={Chengdu}, postcode={610225}, country={China}}
\affiliation[inst2]{organization={School of Statistics, Chengdu University of Information Technology}, city={Chengdu}, postcode={610103}, country={China}}
\affiliation[inst3]{organization={College of Applied Mathematics, Chengdu University of Information Technology}, city={Chengdu}, postcode={610225}, country={China}}
\affiliation[inst4]{organization={School of Artificial Intelligence, Beihang University}, city={Beijing}, postcode={100191}, country={China}}
\affiliation[inst5]{organization={Key Laboratory of Mathematical Meteorology, Chengdu University of Information Technology}, city={Chengdu}, postcode={610225}, country={China}}

\begin{document}

\begin{frontmatter}

\begin{abstract}
% GraphRAG enhances long-document question answering with graph structures, but existing systems face difficult trade-offs among answer quality, computational cost, and knowledge-boundary reliability. Moreover, graphs constructed from long documents are lossy projections of the source text. When retrieved evidence is relevant but insufficient, large language models may still synthesize fluent but unsupported answers. To address these challenges, we propose PAGE-RAG, a Projection-Aware Adaptive Graph-Enhanced RAG framework. Rather than replacing vector retrieval, PAGE-RAG treats graph structures as high-level abstractions of the text and uses governed graph neighborhoods, semantic paths, and community summaries as structural augmentations. Adaptive routing activates graph retrieval and reranking on demand under bounded evidence budgets, while evidence-bounded generation turns evidence sufficiency into an explicit answer-or-abstain decision. Experiments on two English book-length datasets and UltraDomain-Mix show that PAGE-RAG maintains competitive answer quality, lowers query cost, and achieves reliable abstention on unanswerable questions. Ablation results further show that removing the evidence constraint improves apparent accuracy but sharply degrades refusal reliability. Overall, PAGE-RAG adds a more balanced empirical Pareto point across quality and efficiency in long-document GraphRAG.

GraphRAG improves long-document question answering by introducing structured representations beyond conventional retrieval. However, automatically constructed graphs are inherently incomplete projections of source documents, and treating them as independent knowledge sources may lead to unreliable retrieval and generation.
We propose PAGE-RAG, a projection-aware adaptive graph retrieval framework for reliable long-document question answering. PAGE-RAG views graph structures as semantic skeletons that organize and navigate document knowledge, rather than replacing the original knowledge source. Based on this perspective, PAGE-RAG introduces a task-adaptive retrieval routing strategy that dynamically selects appropriate retrieval behaviors according to query requirements. Furthermore, PAGE-RAG incorporates strict knowledge boundary control, ensuring that generated responses remain grounded within available evidence and abstaining from unsupported information beyond the accessible knowledge scope.
Experiments demonstrate that PAGE-RAG achieves competitive answer quality while improving retrieval efficiency and knowledge reliability, highlighting the importance of projection-aware graph modeling, adaptive retrieval, and explicit knowledge boundary control for trustworthy GraphRAG systems.The source code is publicly available at
\url{https://github.com/CXY0112/PAGE-RAG}.
\end{abstract}

\begin{keyword}
GraphRAG \sep retrieval-augmented generation \sep adaptive retrieval \sep knowledge graphs \sep evidence-bounded generation
\end{keyword}

\end{frontmatter}

% Uncomment the following to link to your code, datasets, an extended version or similar.
% You must keep this block between (not within) the abstract and the main body of the paper.
% Make sure that you do not de-anonymize yourself with these links.
% \begin{links}
%     \link{Code}{https://aaai.org/example/code}
%     \link{Datasets}{https://aaai.org/example/datasets}
%     \link{Extended version}{https://aaai.org/example/extended-version}
% \end{links}

\section{Introduction}
Large language models (LLMs) and LLM-based agents have rapidly expanded the range of tasks that can be handled with natural-language interfaces, including question answering, planning, tool use, and interactive knowledge work~\cite{yue2025survey,CHEN2026109006,masterman2024landscape}. As these systems are applied to increasingly complex domains, however, their dependence on finite context windows and parametric knowledge remains a central limitation~\cite{liu2024lost}. Even with longer context models, it is often impractical to place an entire book-length corpus, evolving knowledge base, or domain archive into the prompt for every query~\cite{press2021train,chen2023extending}. More importantly, domain-specific answers require not only broad language competence but also faithful access to the relevant source evidence. When the required evidence is absent, incomplete, or buried in long contexts, LLMs may still produce fluent answers by relying on parametric associations, leading to hallucinations and unverifiable claims~\cite{xu2024knowledge}.

Retrieval-augmented generation (RAG) has become one of the main approaches for mitigating this problem~\cite{lewis2020retrieval}. By retrieving external evidence and conditioning generation on the retrieved context, RAG shifts part of the burden from model parameters to explicit source material. Sparse, dense, and hybrid retrieval methods, often combined with reranking, have shown strong effectiveness for grounding answers in relevant passages~\cite{nogueira2019passage,sun2023chatgpt}. For long documents and domain corpora, RAG provides a practical way to access information that cannot be fully carried in the model context. Nevertheless, standard passage-based RAG primarily treats the corpus as a collection of retrievable text chunks~\cite{gao2023retrieval,barnett2024seven}. It can recover local evidence well, but it provides limited explicit support for relations that span distant passages, multi-hop connections among entities and concepts, and global organization over the document~\cite{he2024g,yao2023editing}.

GraphRAG extends RAG with graph-structured corpus representations, including entities, relations, semantic paths, and communities~\cite{han2025rag}. This family of methods is attractive because graphs can make cross-passage structure explicit and can support relational reasoning or global synthesis that is difficult for passage retrieval alone\citep{edge2024graphrag}. Recent GraphRAG systems have explored community summaries~\cite{traag2019louvain}, graph-vector retrieval~\cite{song2025graph}, path-based reasoning~\cite{zhu2023net}, and other forms of structural augmentation. However, current GraphRAG systems also expose several limitations. Some systems incur high construction or query costs, especially when global reasoning requires broad traversal or community-level map-reduce~\cite{tao2026tagrag}. Some graph representations are noisy or incomplete because they are automatically extracted from long, heterogeneous texts~\cite{chen2026pathrag}. Others may underperform strong passage-retrieval baselines on local factual questions, suggesting that graph structure is not a universal substitute for source passages~\cite{peng2025graph,zhang2025survey}. Finally, many systems rely on prompting the generator to stay grounded, but do not make the evidence boundary an explicit answer-or-refuse decision. As a result, when graph or passage evidence is relevant but insufficient, the model may still synthesize an unsupported answer~\cite{shin2026reasoning}.

We address these limitations from a projection-aware view of GraphRAG. Our hypothesis is that a graph built from a long document should be treated as a useful but lossy projection of the source text, rather than as a faithful replacement for textual evidence. A graph can organize salient entities, relations, events, claims, and communities, but it inevitably omits some atomic facts, qualifications, and local textual details. 
% Therefore, reliable long-document GraphRAG should preserve a textual retrieval floor, use graph computation as an adaptive structural augmentation, and enforce evidence sufficiency at generation time.
Therefore, reliable long-document GraphRAG should optimize not only answer quality and computational efficiency, but also whether the system respects the boundary of available evidence. A system that answers more questions at lower cost is not necessarily preferable if it achieves these gains by producing unsupported answers when evidence is insufficient.

Based on this hypothesis, we propose PAGE-RAG, a \textbf{P}rojection-aware \textbf{A}daptive \textbf{G}raph-\textbf{E}nhanced \textbf{R}etrieval-\textbf{A}ugmented
\textbf{G}eneration framework for long-document question answering. PAGE-RAG maintains both a textual passage index and a governed graph view of the corpus. At query time, it adaptively routes questions to local or global retrieval modes,
activates graph paths and reranking only when needed, and keeps each retrieval branch within a bounded evidence budget. At generation time, PAGE-RAG connects evidence sufficiency to the final answer action: the system answers when the retrieved evidence is sufficient and refuses when it is not. This design aims to improve the empirical trade-off among answer quality, query efficiency, construction cost, and knowledge-boundary reliability, rather than universal dominance over all GraphRAG operating points.

This paper makes the following contributions:
\begin{itemize}
    \item We formulate long-document GraphRAG as a projection-aware retrieval problem, arguing that graph structures provide a compressed semantic skeleton for navigating textual evidence. This motivates an always-on textual retrieval floor guided by governed graph abstractions.
    \item We introduce a query-adaptive retrieval routing strategy that matches retrieval operators to query needs, selecting among textual evidence, graph neighborhoods, semantic paths, community summaries, and reranking. This task-aware routing improves retrieval relevance while bounding graph computation and avoiding unnecessary global expansion.
    % \item We propose evidence-bounded generation, which turns evidence sufficiency into an executable answer boundary. Across two book-length datasets, this mechanism supports reliable abstention on unanswerable questions while maintaining competitive answer quality; on UltraDomain-Mix, PAGE-RAG adds a more balanced empirical operating point between strong passage retrieval and high-cost GraphRAG.
    \item We introduce evidence-bounded generation as an explicit answer-or-abstain mechanism, making boundary reliability a first-class optimization objective alongside answer quality and efficiency. Across two book-length datasets, PAGE-RAG correctly refuses all 24 unanswerable questions; removing the evidence constraint reduces correct refusal to 4/24 despite increasing apparent answer accuracy.
\end{itemize}

% The rest of this paper is organized as follows. We first review related work on neuro-symbolic grounding, RAG, and GraphRAG. We then formalize the long-document GraphRAG setting and present PAGE-RAG's projection-aware hybrid retrieval, adaptive graph computation, and evidence-bounded generation. Finally, we evaluate PAGE-RAG against strong passage and GraphRAG baselines, analyze targeted ablations, and discuss the resulting quality--efficiency--boundary trade-offs.

\section{Related Work}
LLMs store substantial factual and procedural knowledge in their parameters, but this knowledge is difficult to inspect, update, or constrain at inference time~\cite{petroni2019language,meng2022locating,bai2026inference}. Neuro-symbolic methods connect neural models with external symbolic structures or executors to make reasoning more checkable and controllable~\cite{mao2019neuro,hitzler2022neuro,chen2022program,aglionby2022faithful}. For example, PAL translates natural-language problems into executable programs \citep{gao2023pal}, while Logic-LM translates reasoning problems into symbolic formulations solved by deterministic solvers \citep{pan2023logiclm}. RAG can be viewed as the knowledge-access counterpart of this broader grounding paradigm: generation is conditioned on external evidence that can be updated, attributed, and inspected.

Early RAG work combined parametric language models with non-parametric retrieval memories for knowledge-intensive generation \citep{lewis2020rag}. Subsequent work established sparse, dense, hybrid, and reranked retrieval as the main components of practical RAG systems. DPR demonstrated the value of dense passage retrieval for open-domain QA \citep{karpukhin2020dpr}, while BEIR highlighted the complementary cross-domain strengths of lexical, dense, and reranking methods \citep{thakur2021beir}. Passage-based RAG preserves source evidence and supports relatively efficient retrieval, but generally represents a corpus as a flat collection of chunks, providing limited access to cross-passage relations, long-range dependencies, and global thematic structure. RAPTOR addresses part of this limitation by recursively clustering and summarizing chunks into a tree for multi-level retrieval \citep{sarthi2024raptor}. However, hierarchical summaries do not directly expose relation chains or graph topology among entities, events, and concepts.

GraphRAG extends structured retrieval with graph-based corpus representations. Microsoft GraphRAG constructs entity graphs, detects communities, and generates community summaries for query-focused summarization over private corpora \citep{edge2024graphrag}. Its key contribution is to support global sensemaking questions that cannot be answered reliably through local passage retrieval alone. Community-level map-reduce improves coverage for such questions, but can incur substantial construction and query costs when many summaries must be read and synthesized. LightRAG targets these cost and update challenges by combining graph structures with vector representations in a dual-level retrieval scheme \citep{guo2024lightrag}. It retrieves from low-level entity relations and higher-level semantic structures, improving graph-vector retrieval efficiency and incremental updateability. Nevertheless, automatically extracted graphs remain lossy projections of the source text; treating them as replacements for passages may omit local facts, qualifications, and fine-grained constraints.

Other work treats graphs as long-term memory or professional-domain knowledge infrastructure. HippoRAG uses an LLM-built knowledge graph and Personalized PageRank to support multi-hop retrieval and knowledge integration \citep{gutierrez2024hipporag}, while KAG combines knowledge graphs, vector retrieval, and logical-form-guided reasoning for professional-domain QA \citep{liang2024kag}. Together, these systems demonstrate that graphs can serve as memory indexes, semantic navigation layers, reasoning paths, and knowledge-organization structures. Their primary focus, however, is graph retrieval or graph reasoning, leaving less attention to graph projection loss and to the choice of retrieval operators for different query needs.

Overall, the GraphRAG line has progressed from flat passage retrieval to hierarchical summaries, community representations, graph-vector retrieval, and graph-based reasoning. Existing systems address global synthesis, relation modeling, multi-hop retrieval, or professional knowledge organization, but do not yet jointly resolve three issues: graph structures cannot fully replace textual evidence, retrieval computation should adapt to query requirements, and evidence sufficiency should define an explicit answer boundary rather than remain only a prompt-level instruction. These gaps motivate the design of PAGE-RAG.

\section{PAGE-RAG}
PAGE-RAG follows the build--retrieve--generate pipeline. In the build stage, it constructs a hybrid repository that preserves source passages while deriving a governed graph view as a compressed semantic skeleton. In the retrieval stage, it routes each query to retrieval operators that match the query's evidence needs, combining textual evidence, graph neighborhoods, semantic paths, community summaries, and reranking under a bounded evidence budget. In the generation stage, it turns evidence sufficiency into an explicit answer-or-refuse decision. These three stages correspond to the three challenges identified above: graph projection loss, query mismatch across task types, and unreliable knowledge boundaries.

\begin{figure*}[t]
\centering
\makebox[\textwidth][c]{\includegraphics[width=1.15\textwidth]{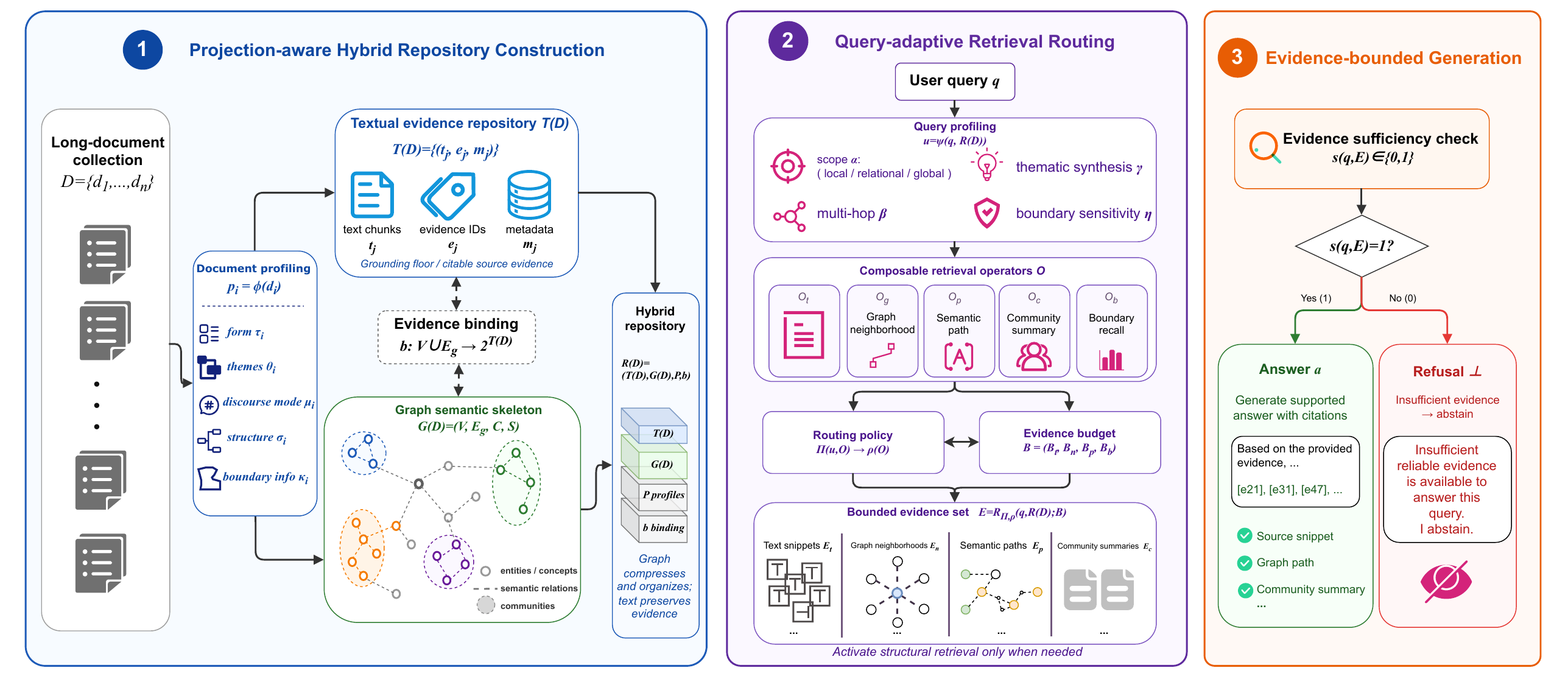}}
\caption{PAGE-RAG framework. The build stage constructs a hybrid repository with both textual evidence and a graph-based semantic skeleton. The retrieval stage routes queries to textual evidence, graph neighborhoods, semantic paths, and community summaries according to query type. The generation stage turns evidence sufficiency into an explicit answer-or-refuse decision.}
\label{fig:framework}
\end{figure*}

\subsection{Projection-Aware Hybrid Repository Construction}
The build stage of PAGE-RAG does not construct only a vector store or only a knowledge graph. Its goal is to construct a hybrid RAG repository that contains both source evidence and a graph-based semantic skeleton. Given a long-document collection $D=\{d_1,\ldots,d_n\}$, PAGE-RAG first profiles each document to determine how it should be segmented, extracted, and organized:
\[
p_i = \Phi(d_i),
\]
where $\Phi$ is the document-profiling function. The profile $p_i$ contains multiple metadata dimensions, which we write as
\[
p_i=(\tau_i,\delta_i,\mu_i,\sigma_i,\kappa_i).
\]
Here, $\tau_i$ denotes the document form, such as fiction, research paper, report, or mixed text; $\delta_i$ denotes the domain or topical distribution; $\mu_i$ denotes the dominant discourse mode, such as narrative, conceptual, or argumentative; $\sigma_i$ denotes internal structure, such as chapters, paragraphs, or topical sections; and $\kappa_i$ denotes boundary information, such as chapter order, timeline, or visible scope. The document profile is not answer evidence by itself. Instead, it acts as a control signal for repository construction and later retrieval.

The document profile first affects the construction of the textual evidence repository $T(D)$. Narrative documents require the system to preserve chapter order, event spans, and character mentions; conceptual or argumentative documents require definitions, propositions, argument chains, and thematic sections to remain recoverable; and global sensemaking corpora require cross-document topics and summary structure. We represent the textual evidence repository as
\[
T(D)=\{(x_j,e_j,m_j)\}_{j=1}^{N},
\]
where $x_j$ is a text chunk or evidence span, $e_j$ is a citable evidence identifier, and $m_j$ is metadata inherited from or derived from document profiles, such as chapter, discourse type, entity coverage, topic labels, and boundary scope. The role of $T(D)$ is to preserve source evidence and provide the grounding floor for all subsequent answers.

The document profile also controls the construction of the graph view $G(D)$. PAGE-RAG does not apply a single fixed schema to all documents. Instead, it selects knowledge units and extraction behavior according to the profile. For a span $x_j$ from document $d_i$, we write this profile-conditioned extraction as
\[
z_j = A_{\mu_i}(x_j,p_i),
\]
where $A_{\mu_i}$ is an extractor controlled by the discourse mode $\mu_i$, and $z_j$ is a candidate knowledge unit. For narrative text, the extracted units may emphasize characters, events, locations, and causal relations. For conceptual text, they may emphasize concepts, definitions, attributes, and theoretical relations. For argumentative text, they may emphasize claims, premises, evidence, and conclusions. This profile-conditioned extraction allows the graph to adapt to heterogeneous long documents rather than forcing all texts into one entity-relation schema.

Candidate knowledge units are then refined into a graph-based semantic skeleton:
\[
G(D)=(V,E_g,C,S),
\]
where $V$ denotes entity, concept, event, or claim nodes; $E_g$ denotes semantic relations or path edges; $C$ denotes community structure; and $S$ denotes community summaries or high-level thematic summaries. We write the governed graph construction process as
\[
G(D)=\Gamma(\{z_j\}_{j=1}^{N},T(D),P),
\]
where $P=\{p_i\}_{i=1}^{n}$ is the set of document profiles, and $\Gamma$ denotes the governed graph-construction function. In practice, $\Gamma$ includes evidence binding, schema and domain-range validation, causal protection, confidence routing, and conservative entity alignment. This formulation emphasizes that the graph is not an independent knowledge base generated freely from text; it is a compressed semantic skeleton constructed under both profile constraints and source-evidence constraints.

To prevent the graph from becoming a lossy substitute for source text, PAGE-RAG maintains bindings between graph elements and textual evidence:
\[
b: V \cup E_g \rightarrow 2^{T(D)}.
\]
The binding function $b$ maps graph nodes and graph edges back to the evidence spans that support them. Through this evidence binding, the graph can guide navigation and evidence organization, while final answers can still be traced back to source text or auditable graph-linked evidence. In other words, $G(D)$ provides semantic compression, relation organization, and retrieval navigation, whereas $T(D)$ provides citable and verifiable factual support.

The resulting hybrid RAG repository is
\[
\mathcal{R}(D)=\big(T(D),G(D),P,b\big).
\]
Here, $T(D)$ is the textual evidence floor, $G(D)$ is the graph-based semantic skeleton, $P$ is the document-profile set that controls construction and retrieval, and $b$ is the evidence binding between graph and text. This construction implements the projection-aware principle: graph structures compress, navigate, and organize long documents without replacing source evidence, while document profiles ensure that different types of texts enter the repository through different extraction and organization strategies.

\subsection{Query-Adaptive Retrieval Routing}
After constructing the hybrid repository $\mathcal{R}(D)=(T(D),G(D),P,b)$, PAGE-RAG treats retrieval as a query-adaptive routing problem rather than a single similarity-ranking problem. Given a user query $q$, the system first estimates a query profile:
\[
u = \Psi(q,\mathcal{R}(D)),
\]
where $\Psi$ is the query-profiling function and $u$ describes the evidence needs and reasoning form of the query. We write this profile as
\[
u=(\alpha,\beta,\gamma,\eta),
\]
where $\alpha$ denotes the query scope, such as local, relational, or global; $\beta$ denotes whether the query requires multi-hop or path-like connections; $\gamma$ denotes whether the query requires thematic synthesis or community-level summaries; and $\eta$ denotes boundary sensitivity, namely whether the query may exceed the current corpus or visible scope. This query profile determines which retrieval operators should be activated and how much budget should be allocated to each evidence type.

PAGE-RAG defines a set of composable retrieval operators:
\[
\mathcal{O}=\{O_t,O_n,O_p,O_c,O_r\}.
\]
Here, $O_t$ is the textual retrieval operator, which retrieves BM25 or dense passages from $T(D)$; $O_n$ is the graph-neighborhood operator, which expands a local subgraph around relevant entities, events, or concepts; $O_p$ is the semantic-path operator, which connects multiple query-related nodes; $O_c$ is the community-summary operator, which retrieves community summaries related to global themes; and $O_r$ is the reranking operator, which selects among candidate evidence items. Instead of treating retrieval as a single-choice route, PAGE-RAG selects a compositional retrieval plan
\[
\rho(q)=\Pi(u,\mathcal{O}),
\]
where $\Pi$ is the routing policy and $\rho(q)\subseteq\mathcal{O}$ is the set of retrieval operators activated for the current query. The plan is multi-select rather than mutually exclusive: a complex query may combine textual retrieval, graph neighborhoods, semantic paths, community summaries, and reranking in the same evidence package. The core principle is that textual evidence remains the grounding floor, while graph neighborhoods, paths, and community summaries are activated when they match the query's evidence needs.

Given a retrieval plan $\rho(q)$, PAGE-RAG constructs a bounded evidence set:
\[
E = R_{\rho(q)}(q,\mathcal{R}(D);B).
\]
The budget $B$ can be distributed across evidence types:
\[
B=(B_t,B_n,B_p,B_c),
\]
where $B_t$ controls the passage or token budget for textual evidence, $B_n$ controls the graph-neighborhood expansion scope, $B_p$ controls the number or length of semantic paths, and $B_c$ controls the number of community summaries. This budgeted construction prevents the system from applying full graph traversal or community-wide synthesis to every query, while still allowing structural evidence to be used when the query requires it.

The core motivation of query-adaptive routing is task adaptation and retrieval relevance. A fixed vector-only route may miss cross-passage relations; a fixed graph-only route may lose source details; and a fixed global route may introduce excessive irrelevant context. PAGE-RAG explicitly links query type, retrieval operators, and evidence budgets through $u$, $\rho(q)$, and $B$, allowing the system to select appropriate evidence structures for local factual, relational, multi-hop, and global synthesis questions. Because each query expands only the passages, graph neighborhoods, semantic paths, or community summaries that match its evidence needs, the system improves retrieval focus while reducing unnecessary context expansion and token consumption.

\subsection{Evidence-Bounded Generation}
The bounded evidence set $E$ produced by retrieval is not always sufficient to support an answer. In long-document RAG, the risk is often not that retrieved evidence is entirely irrelevant, but that it is related yet insufficient: it may mention related themes, characters, events, or concepts without containing the key facts needed to answer the query. If the system is forced to answer in this situation, the LLM may fill the gap with parametric knowledge or semantic association, producing a fluent answer unsupported by the current evidence. Long-document RAG therefore needs to make whether to answer part of the generation decision.

Refusal has two roles in PAGE-RAG. First, it maintains evidence-boundary reliability: the system answers only when the current evidence set is sufficient, and otherwise refuses to package related but insufficient evidence as a cited conclusion. This protects citation traceability and reduces hallucinations induced by partially relevant evidence. Second, refusal provides a preliminary mechanism for access-boundary management in private RAG settings. RAG is often applied to professional domains or private corpora, where sensitive information should be strictly controlled. Although this paper bases refusal mainly on evidence sufficiency rather than a full permission system, it follows the principle that a system should answer according to the evidence boundary available to the current user. This provides a foundation for future private deployment, permission tiers, and sensitive-information control.

PAGE-RAG defines generation as an evidence-bounded answer-or-refuse process. Given a query $q$ and bounded evidence set $E$, the system estimates whether the evidence is sufficient:
\[
s(q,E)\in\{0,1\}.
\]
If $s(q,E)=1$, the generator outputs an answer $a$ supported by citations; if $s(q,E)=0$, the system outputs refusal $\bot$:
\[
y =
\left\{
\begin{array}{ll}
a, & \mathrm{if}\ s(q,E)=1,\\
\bot, & \mathrm{if}\ s(q,E)=0.
\end{array}
\right.
\]
Here, $a$ must be supported by passages, graph-linked evidence, semantic paths, or community summaries in $E$, while $\bot$ indicates that the current evidence is insufficient to answer. Abstention is therefore treated as a correct system action under insufficient evidence, rather than as a generation failure.

This mechanism differs from prompt-only grounding. Asking the model to answer from context is a soft instruction that still depends on model compliance, whereas evidence-bounded generation turns evidence sufficiency into an explicit output boundary. The final refusal boundary prevents related but insufficient evidence from being completed into unsupported claims, and provides an interface for future permission tiers and private knowledge governance.

\section{Evaluation}
We evaluate whether PAGE-RAG provides competitive long-document QA quality, improves the balance among quality, query efficiency, and knowledge-boundary reliability, and whether its key components contribute as intended.

\subsection{Experimental Setup}
\textbf{Datasets.} We evaluate on two English book-length datasets and one public global sensemaking benchmark. \emph{Simulacra and Simulation}~\cite{baudrillard1994simulacra} is a conceptual, argumentative book dataset with 94 answerable and 12 unanswerable questions; its abstract concepts and distributed arguments stress conceptual relation recovery beyond entity lookup. \emph{One Hundred Years of Solitude}~\cite{marquez2018one} is a narrative novel with 88 answerable and 12 unanswerable questions; as a relationally complex novel, its character interactions and events form long-range narrative dependencies. These two datasets test local evidence recovery, cross-passage reasoning, thematic synthesis, and closed-corpus refusal. UltraDomain-Mix~\cite{tao2026tagrag} contains 125 public global sensemaking questions over 61 documents and 619K tokens, and is evaluated with reference-free pairwise judgments. This public, heterogeneous benchmark tests generalization beyond single-book corpora to cross-domain global synthesis.

\textbf{Systems.} We compare PAGE-RAG with three baselines. BDR-RAG is our controlled no-graph passage baseline, combining BM25, dense retrieval, and reranking over the same source spans, following established retrieval components in DPR and BEIR \citep{karpukhin2020dpr,thakur2021beir}. It is intentionally strong and incurs no LLM-based graph construction cost. LightRAG~\cite{guo2024lightrag} is a lightweight graph-vector GraphRAG system. Microsoft GraphRAG~\cite{edge2024graphrag} is a heavier community-summary and map-reduce style GraphRAG system. We keep system-specific behavior documented rather than modifying baselines to remove their defining retrieval mechanisms.

\textbf{Configuration and metrics.} All systems use the same source corpora, the same answer model, the same embedding model, and comparable chunking or context settings when supported by the baseline. For the two books, we report strict and lenient accuracy, correct refusal on unanswerable questions, median latency, query tokens per question, and one-time build tokens. For ablations, we use boundary-balanced score (BBS), the harmonic mean of lenient accuracy and correct-refusal rate: $BBS=2\cdot Acc_{lenient}\cdot Ref/(Acc_{lenient}+Ref)$. For UltraDomain-Mix, we report position-debiased pairwise judgments on four dimensions--comprehensiveness, diversity, empowerment, and directness--together with query tokens and latency.

\subsection{Main Results}

\begin{figure*}[h]
\centering
\includegraphics[width=1.0\textwidth]{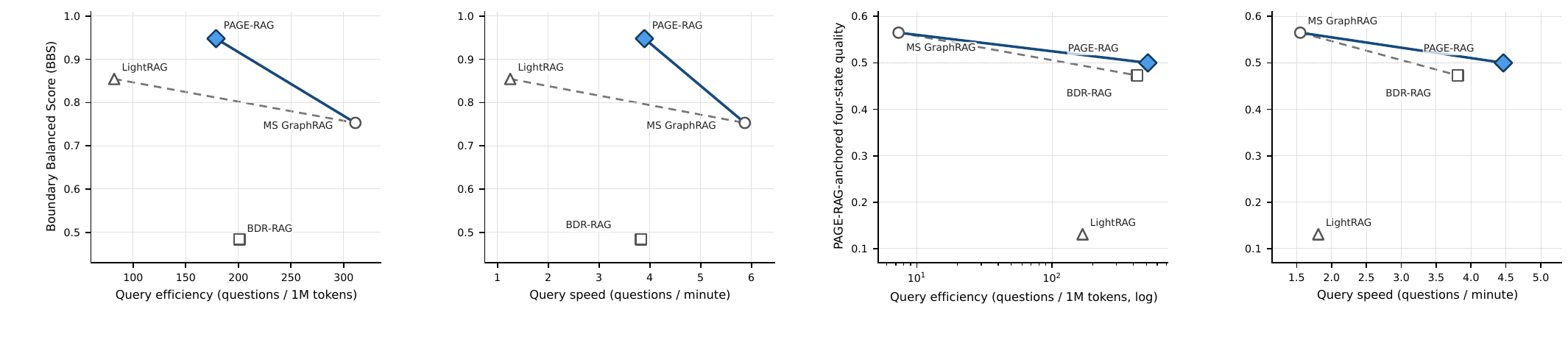}
\caption{Empirical quality--efficiency frontier across book-length QA and UltraDomain-Mix. Dashed segments show the baseline-only frontier and solid segments show the frontier after adding PAGE-RAG. PAGE-RAG adds a balanced operating point rather than universally dominating every baseline.}
\label{fig:pareto-frontier}
\end{figure*}

Figure~\ref{fig:pareto-frontier} visualizes the empirical frontier induced by the main results. For the two book datasets, quality combines lenient accuracy and correct refusal, so the plot reflects both answer correctness and boundary reliability. For UltraDomain-Mix, quality is the PAGE-RAG-anchored four-state pairwise index. The figure shows that PAGE-RAG is not merely a point with lower cost; it occupies a different region of the quality--efficiency space. On the book datasets, it combines the strongest boundary-balanced quality with substantially lower latency and query cost than LightRAG. On UltraDomain-Mix, Microsoft GraphRAG remains the higher-quality extreme, but PAGE-RAG reaches a nearby balanced region with far lower query tokens and latency.

\textbf{Book-length QA.} Table~\ref{tab:book-main} shows the main results on the two book-length datasets. PAGE-RAG is not universally best on strict accuracy, However, strict or lenient accuracy alone does not capture whether a system answers beyond the available evidence. LightRAG leads strict accuracy on both books, and BDR-RAG is very strong on the conceptual book. However, PAGE-RAG is the only system that combines competitive lenient accuracy with stable correct refusal on both datasets. On \emph{Simulacra and Simulation}, PAGE-RAG reaches 92.6\% lenient accuracy and 12/12 correct refusal; on \emph{One Hundred Years of Solitude}, it reaches the highest lenient accuracy, 87.5\%, and again refuses all 12 unanswerable questions correctly. This pattern supports the role of evidence-bounded generation: the system does not merely answer more often, but preserves a closed-corpus boundary.

\begin{table}[h]
\centering
\small
\caption{Main results on the two book-length datasets. Acc. denotes accuracy, Ref. denotes correct refusal, Lat. denotes median latency, QTok denotes query tokens per question, and BTok denotes build LLM tokens.}
\label{tab:book-main}
\resizebox{\columnwidth}{!}{%
\begin{tabular}{llrrrrrr}
\toprule
Dataset & System & Strict & Lenient & Ref. & Lat. & QTok & BTok \\
\midrule
Simulacra & PAGE-RAG & 72.3 & 92.6 & 12/12 & 14.9s & 4,213 & 842K \\
Simulacra & BDR-RAG & 77.7 & 93.6 & 5/12 & 15.9s & 3,808 & 0 \\
Simulacra & LightRAG & 80.9 & 85.1 & 12/12 & 47.3s & 9,233 & 686K \\
Simulacra & MS GraphRAG & 52.1 & 67.0 & 11/12 & 11.8s & 3,281 & 1.30M \\
\midrule
Solitude & PAGE-RAG & 60.2 & 87.5 & 12/12 & 16.3s & 7,052 & 1.85M \\
Solitude & BDR-RAG & 54.5 & 81.8 & 3/12 & 15.4s & 6,215 & 0 \\
Solitude & LightRAG & 64.8 & 81.8 & 9/12 & 48.4s & 15,305 & 1.78M \\
Solitude & MS GraphRAG & 48.9 & 70.5 & 9/12 & 8.5s & 3,142 & 4.41M \\
\bottomrule
\end{tabular}%
}
\end{table}

\begin{table*}[htpb]
\centering
\small
\caption{UltraDomain-Mix four-dimensional pairwise results and efficiency. Each quality cell reports PAGE-RAG wins / ties / baseline wins under position-debiased judgments.}
\label{tab:ultradomain}
\resizebox{\textwidth}{!}{%
\begin{tabular}{lrrrrrr}
\toprule
Comparison & Comp. & Div. & Emp. & Direct. & QTok / q & Lat. \\
\midrule
PAGE-RAG vs BDR-RAG & 49/41/35 & 51/37/37 & 40/48/37 & 23/75/27 & 1,956 vs 2,351 & 13.4s vs 15.75s \\
PAGE-RAG vs MS GraphRAG & 39/21/65 & 36/26/63 & 40/29/56 & 27/75/23 & 1,956 vs 136,509 & 13.4s vs 38.7s \\
PAGE-RAG vs LightRAG & 101/22/2 & 102/20/3 & 103/18/4 & 76/45/4 & 1,956 vs 5,949 & 13.4s vs 33.1s \\
\bottomrule
\end{tabular}%
}
\end{table*}

The two books also show why PAGE-RAG does not treat graph retrieval as a universal replacement for passage retrieval. On the conceptual book, the strong passage baseline is already highly competitive on answerable questions, suggesting that local textual evidence often suffices. On the narrative book, PAGE-RAG obtains the best lenient accuracy, consistent with graph-guided evidence organization being more useful for cross-chapter events, characters, and relations. This supports the query-adaptive design: graph structure is most useful when the question requires structural organization, while the textual floor remains critical for grounding.

\textbf{UltraDomain-Mix.} Table~\ref{tab:ultradomain} reports four-dimensional pairwise results on UltraDomain-Mix. PAGE-RAG is close to BDR-RAG across comprehensiveness and diversity, weaker on directness, and strongly ahead of LightRAG across all four dimensions. Microsoft GraphRAG wins more often on the four quality dimensions, but its cost profile is substantially different: Microsoft GraphRAG uses 136,509 query tokens per question, while PAGE-RAG uses 1,956. This result positions Microsoft GraphRAG as a high-quality, high-cost extreme for this setting, while PAGE-RAG adds a more balanced operating point with lower latency and much lower query cost.

% Taken together, the main results support PAGE-RAG's intended positioning. It is not a universal quality-dominant system, and it does not replace every GraphRAG operating point. Instead, it combines competitive answer quality, reliable abstention on closed-corpus book questions, and low query-time cost, thereby adding a balanced quality--efficiency--boundary point.

Taken together, the main results support PAGE-RAG's intended positioning. It is not a universal quality-dominant system, and it does not replace every GraphRAG operating point. Instead, it combines competitive answer quality, reliable abstention on closed-corpus book questions, and low query-time cost, thereby occupying a balanced empirical position on the quality--efficiency--boundary Pareto frontier.

\subsection{Ablation Study}
Table~\ref{tab:ablation} reports the core ablations on the two book-length datasets. The results isolate two dominant mechanisms. First, the textual retrieval floor is the basis of evidence supply: removing it reduces lenient accuracy from 90.1\% to 58.2\%, even though the system still refuses all unanswerable questions correctly. This supports the projection-aware claim that graph structures can organize and navigate evidence, but should not replace source passages. Second, the evidence constraint is the direct mechanism for the knowledge boundary. 
% When it is relaxed, lenient accuracy rises to 95.1\%, but correct refusal drops from 24/24 to 4/24; manual audit finds substantive out-of-evidence answers on 22 of 24 unanswerable questions. Thus, higher answer rate can improve apparent answerable-question accuracy while collapsing boundary reliability.
Removing the evidence constraint creates an apparently better accuracy point but a substantially worse system operating point. Lenient accuracy rises from 90.1\% to 95.1\%, while correct refusal collapses from 24/24 to 4/24, and 22 of 24 unanswerable questions receive substantive out-of-evidence answers. This result shows that answer accuracy and boundary reliability are not interchangeable: a system may move upward on an accuracy-only evaluation while moving sharply backward on the true quality–efficiency–reliability frontier.

\begin{table}[h]
\centering
\small
\caption{Core ablations on the two book-length datasets. BBS denotes boundary-balanced score; Ref. reports correct refusals on unanswerable questions.}
\label{tab:ablation}
\resizebox{\columnwidth}{!}{%
\begin{tabular}{lccc}
\toprule
Variant & Lenient Acc. & Ref. & BBS  \\
\midrule
Full PAGE-RAG & 90.1 & 24/24 & 94.8 \\
No Floor & 58.2 & 24/24 & 73.6  \\
Raw Graph & 86.3 & 24/24 & 92.6 \\
No Evidence Constraint & 95.1 & 4/24 & 28.4 \\
No Structural Guidance & 87.9 & 24/24 & 93.6  \\
\bottomrule
\end{tabular}%
}
\end{table}

Graph governance and structural guidance mainly contribute organization, retrievability, and targeted gains. Raw Graph obtains lower lenient accuracy than the full system, indicating that unguided automatic graph extraction can introduce redundant entities, noisy relations, and fragmented communities that weaken the graph as a retrieval skeleton. The structural evidence is more direct on Solitude: Raw Graph contains 4,300 entities, 4,827 semantic edges, and 1,024 communities, while the full system compresses them to 1,919 entities, 3,791 semantic edges, and 194 communities. Isolated entities decrease from 763 to 175, and actual path use increases from 3 to 39 queries. On Simulacra, the full system also reduces entities from 1,904 to 1,555 and communities from 484 to 351, while increasing path use from 18 to 32 queries. These results show that graph governance is not merely an auxiliary cleanup step; it reduces entity duplication, community fragmentation, and noisy structural context, making the graph more suitable as a compressed semantic skeleton. A more compact graph lowers the search space and path-enumeration difficulty, reduces the chance that irrelevant structure enters the evidence context, and can therefore improve retrieval quality while helping control token cost and query efficiency.

No Structural Guidance further shows that structural evidence is most useful when activated selectively. Removing explicit paths and structural hints lowers lenient accuracy from 90.1\% to 87.9\%; on routed path-using answerable questions, the drop is larger, from 94.1\% to 89.7\%. This suggests that explicit graph paths help most when the router identifies structural evidence needs, rather than serving as a universal context expansion. Overall, the ablations support a mechanism account: textual retrieval supplies grounded evidence, graph governance makes the semantic skeleton compact and searchable, query-adaptive structural guidance provides targeted gains, and evidence-bounded generation enforces the answer boundary.
\section{Conclusion}
This paper introduced PAGE-RAG to address the trade-off among answer quality, efficiency, and knowledge-boundary reliability in long-document GraphRAG. Its central principle is projection awareness: PAGE-RAG combines an always-on textual retrieval floor with graph structures that serve as compressed semantic skeletons for navigating and organizing source evidence. Through governed graph abstractions, query-adaptive routing, bounded structural retrieval, and evidence-bounded generation, PAGE-RAG adds a balanced empirical operating point to the quality--efficiency frontier while enforcing a strict answer-or-refuse boundary when available evidence is insufficient.

Experiments on two books and UltraDomain-Mix show that PAGE-RAG adds a more balanced empirical operating point rather than dominating every system on every single quality metric. It maintains competitive answer quality, reduces query-time token and latency cost, and preserves reliable abstention on closed-corpus questions. Ablations further show that the textual floor supplies grounded evidence, graph governance improves structural retrievability, and evidence constraints are central to boundary reliability. Future work will explore finer-grained access control, cross-domain graph construction, and more precise task-adaptive graph computation.

\bibliographystyle{elsarticle-num}
\bibliography{elsarticle-template-num}

@inproceedings{gao2023pal,
  title={Pal: Program-aided language models},
  author={Gao, Luyu and Madaan, Aman and Zhou, Shuyan and Alon, Uri and Liu, Pengfei and Yang, Yiming and Callan, Jamie and Neubig, Graham},
  booktitle={International conference on machine learning},
  pages={10764--10799},
  year={2023},
  organization={PMLR}
}

@inproceedings{pan2023logiclm,
  title={Logic-lm: Empowering large language models with symbolic solvers for faithful logical reasoning},
  author={Pan, Liangming and Albalak, Alon and Wang, Xinyi and Wang, William},
  booktitle={Findings of the Association for Computational Linguistics: EMNLP 2023},
  pages={3806--3824},
  year={2023}
}

@inproceedings{lewis2020rag,
 author = {Lewis, Patrick and Perez, Ethan and Piktus, Aleksandra and Petroni, Fabio and Karpukhin, Vladimir and Goyal, Naman and K\"{u}ttler, Heinrich and Lewis, Mike and Yih, Wen-tau and Rockt\"{a}schel, Tim and Riedel, Sebastian and Kiela, Douwe},
 booktitle = {Advances in Neural Information Processing Systems},
 editor = {H. Larochelle and M. Ranzato and R. Hadsell and M.F. Balcan and H. Lin},
 pages = {9459--9474},
 publisher = {Curran Associates, Inc.},
 title = {Retrieval-Augmented Generation for Knowledge-Intensive NLP Tasks},
 url = {https://proceedings.neurips.cc/paper_files/paper/2020/file/6b493230205f780e1bc26945df7481e5-Paper.pdf},
 volume = {33},
 year = {2020}
}

@inproceedings{karpukhin2020dpr,
  title={Dense passage retrieval for open-domain question answering},
  author={Karpukhin, Vladimir and Oguz, Barlas and Min, Sewon and Lewis, Patrick and Wu, Ledell and Edunov, Sergey and Chen, Danqi and Yih, Wen-tau},
  booktitle={Proceedings of the 2020 conference on empirical methods in natural language processing (EMNLP)},
  pages={6769--6781},
  year={2020}
}

@inproceedings{thakur2021beir,
  title = {{BEIR}: A Heterogeneous Benchmark for Zero-shot Evaluation of Information Retrieval Models},
  author = {Thakur, Nandan and Reimers, Nils and R{\"u}ckl{\'e}, Andreas and Srivastava, Abhishek and Gurevych, Iryna},
  booktitle = {Thirty-fifth Conference on Neural Information Processing Systems Datasets and Benchmarks Track},
  year = {2021}
}

@inproceedings{sarthi2024raptor,
  title={Raptor: Recursive abstractive processing for tree-organized retrieval},
  author={Sarthi, Parth and Abdullah, Salman and Tuli, Aditi and Khanna, Shubh and Goldie, Anna and Manning, Christopher},
  booktitle={International Conference on Learning Representations},
  volume={2024},
  pages={32628--32649},
  year={2024}
}

@misc{edge2024graphrag,
  title = {From Local to Global: A Graph {RAG} Approach to Query-Focused Summarization},
  author = {Edge, Darren and Trinh, Ha and Cheng, Newman and Bradley, Joshua and Chao, Alex and Mody, Apurva and Truitt, Steven and Larson, Jonathan},
  year = {2024},
  eprint = {2404.16130},
  archivePrefix = {arXiv},
  primaryClass = {cs.CL}
}

@inproceedings{guo2024lightrag,
    title = "{L}ight{RAG}: Simple and Fast Retrieval-Augmented Generation",
    author = "Guo, Zirui  and
      Xia, Lianghao  and
      Yu, Yanhua  and
      Ao, Tu  and
      Huang, Chao",
    editor = "Christodoulopoulos, Christos  and
      Chakraborty, Tanmoy  and
      Rose, Carolyn  and
      Peng, Violet",
    booktitle = "Findings of the Association for Computational Linguistics: EMNLP 2025",
    month = nov,
    year = "2025",
    address = "Suzhou, China",
    publisher = "Association for Computational Linguistics",
    url = "https://aclanthology.org/2025.findings-emnlp.568/",
    doi = "10.18653/v1/2025.findings-emnlp.568",
    pages = "10746--10761",
    ISBN = "979-8-89176-335-7",
}

@inproceedings{liang2024kag,
  title={Kag: Boosting llms in professional domains via knowledge augmented generation},
  author={Liang, Lei and Bo, Zhongpu and Gui, Zhengke and Zhu, Zhongshu and Zhong, Ling and Zhao, Peilong and Sun, Mengshu and Zhang, Zhiqiang and Zhou, Jun and Chen, Wenguang and others},
  booktitle={Companion Proceedings of the ACM on Web Conference 2025},
  pages={334--343},
  year={2025}
}

@article{gutierrez2024hipporag,
  title={Hipporag: Neurobiologically inspired long-term memory for large language models},
  author={Guti{\'e}rrez, Bernal J and Shu, Yiheng and Gu, Yu and Yasunaga, Michihiro and Su, Yu},
  journal={Advances in neural information processing systems},
  volume={37},
  pages={59532--59569},
  year={2024}
}

@article{yue2025survey,
  title={A survey of large language model agents for question answering},
  author={Yue, Murong},
  journal={arXiv preprint arXiv:2503.19213},
  year={2025}
}

@article{CHEN2026109006,
title = {KG -augmented executable CoT for mathematical coding},
journal = {Neural Networks},
volume = {202},
pages = {109006},
year = {2026},
issn = {0893-6080},
doi = {https://doi.org/10.1016/j.neunet.2026.109006},
url = {https://www.sciencedirect.com/science/article/pii/S0893608026004673},
author = {Xingyu Chen and Junxiu An and Jun Guo and Li Wang and Jingcai Guo},
}

@article{masterman2024landscape,
  title={The landscape of emerging ai agent architectures for reasoning, planning, and tool calling: A survey},
  author={Masterman, Tula and Besen, Sandi and Sawtell, Mason and Chao, Alex},
  journal={arXiv preprint arXiv:2404.11584},
  year={2024}
}

@article{liu2024lost,
  title={Lost in the middle: How language models use long contexts},
  author={Liu, Nelson F and Lin, Kevin and Hewitt, John and Paranjape, Ashwin and Bevilacqua, Michele and Petroni, Fabio and Liang, Percy},
  journal={Transactions of the association for computational linguistics},
  volume={12},
  pages={157--173},
  year={2024}
}

@article{press2021train,
  title={Train short, test long: Attention with linear biases enables input length extrapolation},
  author={Press, Ofir and Smith, Noah A and Lewis, Mike},
  journal={arXiv preprint arXiv:2108.12409},
  year={2021}
}

@article{chen2023extending,
  title={Extending context window of large language models via positional interpolation},
  author={Chen, Shouyuan and Wong, Sherman and Chen, Liangjian and Tian, Yuandong},
  journal={arXiv preprint arXiv:2306.15595},
  year={2023}
}

@inproceedings{xu2024knowledge,
  title={Knowledge conflicts for llms: A survey},
  author={Xu, Rongwu and Qi, Zehan and Guo, Zhijiang and Wang, Cunxiang and Wang, Hongru and Zhang, Yue and Xu, Wei},
  booktitle={Proceedings of the 2024 Conference on Empirical Methods in Natural Language Processing},
  pages={8541--8565},
  year={2024}
}

@article{lewis2020retrieval,
  title={Retrieval-augmented generation for knowledge-intensive nlp tasks},
  author={Lewis, Patrick and Perez, Ethan and Piktus, Aleksandra and Petroni, Fabio and Karpukhin, Vladimir and Goyal, Naman and K{\"u}ttler, Heinrich and Lewis, Mike and Yih, Wen-tau and Rockt{\"a}schel, Tim and others},
  journal={Advances in neural information processing systems},
  volume={33},
  pages={9459--9474},
  year={2020}
}

@article{nogueira2019passage,
  title={Passage Re-ranking with BERT},
  author={Nogueira, Rodrigo and Cho, Kyunghyun},
  journal={arXiv preprint arXiv:1901.04085},
  year={2019}
}

@inproceedings{sun2023chatgpt,
  title={Is ChatGPT good at search? investigating large language models as re-ranking agents},
  author={Sun, Weiwei and Yan, Lingyong and Ma, Xinyu and Wang, Shuaiqiang and Ren, Pengjie and Chen, Zhumin and Yin, Dawei and Ren, Zhaochun},
  booktitle={Proceedings of the 2023 conference on empirical methods in natural language processing},
  pages={14918--14937},
  year={2023}
}

@article{he2024g,
  title={G-retriever: Retrieval-augmented generation for textual graph understanding and question answering},
  author={He, Xiaoxin and Tian, Yijun and Sun, Yifei and Chawla, Nitesh V and Laurent, Thomas and LeCun, Yann and Bresson, Xavier and Hooi, Bryan},
  journal={Advances in Neural Information Processing Systems},
  volume={37},
  pages={132876--132907},
  year={2024}
}

@article{han2025rag,
  title={Rag vs. graphrag: A systematic evaluation and key insights},
  author={Han, Haoyu and Ma, Li and Wang, Yu and Shomer, Harry and Lei, Yongjia and Qi, Zhisheng and Guo, Kai and Hua, Zhigang and Long, Bo and Liu, Hui and others},
  journal={arXiv preprint arXiv:2502.11371},
  year={2025}
}

@article{peng2025graph,
  title={Graph retrieval-augmented generation: A survey},
  author={Peng, Boci and Zhu, Yun and Liu, Yongchao and Bo, Xiaohe and Shi, Haizhou and Hong, Chuntao and Zhang, Yan and Tang, Siliang},
  journal={ACM Transactions on Information Systems},
  volume={44},
  number={2},
  pages={1--52},
  year={2025},
  publisher={ACM New York, NY}
}

@article{zhang2025survey,
  title={A survey of graph retrieval-augmented generation for customized large language models},
  author={Zhang, Qinggang and Chen, Shengyuan and Bei, Yuanchen and Yuan, Zheng and Zhou, Huachi and Hong, Zijin and Chen, Hao and Xiao, Yilin and Zhou, Chuang and Dong, Junnan and others},
  journal={arXiv preprint arXiv:2501.13958},
  year={2025}
}

@article{shin2026reasoning,
  title={The Reasoning Trap: An Information-Theoretic Bound on Closed-System Multi-Step LLM Reasoning},
  author={Shin, Kwan Soo},
  journal={arXiv preprint arXiv:2605.01704},
  year={2026}
}

@inproceedings{tao2026tagrag,
  title={TagRAG: Tag-guided Hierarchical Knowledge Graph Retrieval-Augmented Generation},
  author={Tao, Wenbiao and Li, Xinyuan and Lan, Yunshi and Qian, Weining},
  booktitle={Findings of the Association for Computational Linguistics: ACL 2026},
  pages={6434--6456},
  year={2026}
}

@inproceedings{yao2023editing,
  title={Editing large language models: Problems, methods, and opportunities},
  author={Yao, Yunzhi and Wang, Peng and Tian, Bozhong and Cheng, Siyuan and Li, Zhoubo and Deng, Shumin and Chen, Huajun and Zhang, Ningyu},
  booktitle={Proceedings of the 2023 Conference on Empirical Methods in Natural Language Processing},
  pages={10222--10240},
  year={2023}
}

@incollection{marquez2018one,
  title={One hundred years of solitude},
  author={M{\'a}rquez, Gabriel Garc{\'\i}a},
  booktitle={Medicine and literature, volume two},
  pages={255--272},
  year={2018},
  publisher={CRC Press}
}

@book{baudrillard1994simulacra,
  title={Simulacra and simulation},
  author={Baudrillard, Jean},
  year={1994},
  publisher={University of Michigan press}
}

@inproceedings{petroni2019language,
  title={Language models as knowledge bases?},
  author={Petroni, Fabio and Rockt{\"a}schel, Tim and Riedel, Sebastian and Lewis, Patrick and Bakhtin, Anton and Wu, Yuxiang and Miller, Alexander},
  booktitle={Proceedings of the 2019 conference on empirical methods in natural language processing and the 9th international joint conference on natural language processing (EMNLP-IJCNLP)},
  pages={2463--2473},
  year={2019}
}

@article{meng2022locating,
  title={Locating and editing factual associations in gpt},
  author={Meng, Kevin and Bau, David and Andonian, Alex and Belinkov, Yonatan},
  journal={Advances in neural information processing systems},
  volume={35},
  pages={17359--17372},
  year={2022}
}

@article{bai2026inference,
  title={Inference-Time Control for Trustworthy Large Language Models},
  author={Bai, Yuyang and Liu, Zheyuan and Yan, Han and Xu, Zhangchen and Wan, Yixin and Chen, Canyu and Wang, Zehong and Yuan, Xiangchi and Huang, Yue and Dou, Guangyao and others},
  year={2026},
  publisher={Preprints}
}

@article{mao2019neuro,
  title={The neuro-symbolic concept learner: Interpreting scenes, words, and sentences from natural supervision},
  author={Mao, Jiayuan and Gan, Chuang and Kohli, Pushmeet and Tenenbaum, Joshua B and Wu, Jiajun},
  journal={arXiv preprint arXiv:1904.12584},
  year={2019}
}

@article{hitzler2022neuro,
  title={Neuro-symbolic approaches in artificial intelligence},
  author={Hitzler, Pascal and Eberhart, Aaron and Ebrahimi, Monireh and Sarker, Md Kamruzzaman and Zhou, Lu},
  journal={National Science Review},
  volume={9},
  number={6},
  pages={nwac035},
  year={2022},
  publisher={Oxford University Press}
}

@article{chen2022program,
  title={Program of thoughts prompting: Disentangling computation from reasoning for numerical reasoning tasks},
  author={Chen, Wenhu and Ma, Xueguang and Wang, Xinyi and Cohen, William W},
  journal={arXiv preprint arXiv:2211.12588},
  year={2022}
}

@inproceedings{aglionby2022faithful,
  title={Faithful knowledge graph explanations in commonsense question answering},
  author={Aglionby, Guy and Teufel, Simone},
  booktitle={Proceedings of the 2022 Conference on Empirical Methods in Natural Language Processing},
  pages={10811--10817},
  year={2022}
}

@inproceedings{chen2026pathrag,
  title={Pathrag: Pruning graph-based retrieval augmented generation with relational paths},
  author={Chen, Boyu and Guo, Zirui and Yang, Zidan and Chen, Yuluo and Chen, Junze and Liu, Zhenghao and Shi, Chuan and Yang, Cheng},
  booktitle={Proceedings of the AAAI conference on artificial intelligence},
  volume={40},
  number={36},
  pages={30183--30191},
  year={2026}
}

@article{traag2019louvain,
  title={From Louvain to Leiden: guaranteeing well-connected communities},
  author={Traag, Vincent A and Waltman, Ludo and Van Eck, Nees Jan},
  journal={Scientific reports},
  volume={9},
  number={1},
  pages={5233},
  year={2019},
  publisher={Nature Publishing Group UK London}
}

@article{song2025graph,
  title={Graph retrieval augmented large language models for facial phenotype associated rare genetic disease},
  author={Song, Jie and Xu, Zhichuan and He, Mengqiao and Feng, Jinhua and Shen, Bairong},
  journal={NPJ digital medicine},
  volume={8},
  number={1},
  pages={543},
  year={2025},
  publisher={Nature Publishing Group UK London}
}

@article{zhu2023net,
  title={A* net: A scalable path-based reasoning approach for knowledge graphs},
  author={Zhu, Zhaocheng and Yuan, Xinyu and Galkin, Michael and Xhonneux, Louis-Pascal and Zhang, Ming and Gazeau, Maxime and Tang, Jian},
  journal={Advances in neural information processing systems},
  volume={36},
  pages={59323--59336},
  year={2023}
}

@article{gao2023retrieval,
  title={Retrieval-augmented generation for large language models: A survey},
  author={Gao, Yunfan and Xiong, Yun and Gao, Xinyu and Jia, Kangxiang and Pan, Jinliu and Bi, Yuxi and Dai, Yi and Sun, Jiawei and Wang, Meng and Wang, Haofen},
  journal={arXiv preprint arXiv:2312.10997},
  year={2023}
}

@inproceedings{barnett2024seven,
  title={Seven failure points when engineering a retrieval augmented generation system},
  author={Barnett, Scott and Kurniawan, Stefanus and Thudumu, Srikanth and Brannelly, Zach and Abdelrazek, Mohamed},
  booktitle={Proceedings of the IEEE/ACM 3rd International Conference on AI Engineering-Software Engineering for AI},
  pages={194--199},
  year={2024}
}

% Check whether the conference requires a reproducibility checklist to be included in the paper.
% If so, you can uncomment the following line and ajust the path to include it.
% \input{ReproducibilityChecklist.tex}

\end{document}